\documentclass[journal]{IEEEtran}
\usepackage{amsmath,amssymb,amsfonts}
\usepackage{algorithmic}
\usepackage{algorithm}
\usepackage{array}
\usepackage[caption=false,font=normalsize,labelfont=sf,textfont=sf]{subfig}
\usepackage{textcomp}
\usepackage{stfloats}
\usepackage{url}
\usepackage{verbatim}
\usepackage{graphicx}
\usepackage{cite}
\usepackage{color}

\newtheorem{theorem}{Theorem}
\newtheorem{corollary}{Corollary}

\makeatletter
\def\endthebibliography{%
  \def\@noitemerr{\@latex@warning{Empty `thebibliography' environment}}%
  \endlist
}
\makeatother

\begin{document}

\title{On Ambiguity in Linear Inverse Problems: Entrywise Bounds on Nearly Data-Consistent Solutions and Entrywise Condition Numbers}

\author{Justin P. Haldar,~\IEEEmembership{Senior Member,~IEEE,}%
\thanks{This work was supported in part by NIH research grants R01-MH116173 and  R01-NS074980.}%
\thanks{J. Haldar is with the Signal and Image Processing Institute, Ming Hsieh Department of Electrical and Computer Engineering, University of Southern California, Los Angeles, CA, 90089 USA (e-mail: jhaldar@usc.edu).}}

\maketitle

\begin{abstract}
Ill-posed linear inverse problems appear frequently in various signal processing applications. It can be very useful to have theoretical characterizations that quantify the level of ill-posedness for a given inverse problem and the degree of ambiguity that may exist about its solution.   Traditional measures of ill-posedness, such as the condition number of a matrix, provide characterizations that are global in nature. While such characterizations can be powerful, they can also fail to provide full insight into situations where certain entries of the solution vector are more or less ambiguous than others.  In this work, we derive novel theoretical lower- and upper-bounds that apply to individual entries of the solution vector, and are valid for all potential solution vectors that are nearly data-consistent.  These bounds are agnostic to the noise statistics  and the specific method used to solve the inverse problem, and are also shown to be tight.  In addition, our results also lead us to introduce an entrywise version of the traditional condition number, which provides a substantially more nuanced characterization of scenarios where certain elements of the solution vector are less sensitive to perturbations than others.  Our results are illustrated in an application to magnetic resonance imaging reconstruction, and we include discussions of practical computation methods for large-scale inverse problems, connections between our new theory and the traditional Cram\'{e}r-Rao bound under statistical modeling assumptions, and potential extensions to cases involving constraints beyond just data-consistency.
\end{abstract}


\section{Introduction}
This paper concerns the classical and well-studied finite-dimensional linear inverse problem of estimating a true unknown signal vector $\mathbf{x}^* \in \mathbb{R}^N$ from a noisy data vector $\mathbf{b} \in \mathbb{R}^M$ obtained as
\begin{equation}
 \mathbf{b} = \mathbf{A}\mathbf{x}^* + \mathbf{n},\label{eq:model}
\end{equation}
where $\mathbf{A} \in \mathbb{R}^{M\times N}$ is the known system matrix and $\mathbf{n} \in \mathbb{R}^M$ represents an unknown noise perturbation. In many practical scenarios, this inverse problem is ill-posed, meaning that the original value of $\mathbf{x}^*$ might be ambiguous even in the absence of noise, and/or that solutions may be highly sensitive to  noise.  

While there are many existing theoretical results that characterize ambiguity for this inverse problem, it is frequent to see characterizations phrased in terms of global bounds on the estimation error as a function of the size of the noise \cite{golub1996,bertero1998, moon2000,heath2002,bresler2008a}.  To be concrete, if we let $\hat{\mathbf{x}}$ be the estimate of $\mathbf{x}^*$ obtained through some estimation procedure, it is common to see global error bounds of the form
\begin{equation}
 \|\hat{\mathbf{x}} - \mathbf{x}^*\|_2 \leq C \|\mathbf{n}\|_2,\label{eq:global}
\end{equation}
where $\|\cdot\|_2$ denotes the standard euclidean norm and the constant $C$  will depend on the characteristics of the estimation procedure, the characteristics of the matrix $\mathbf{A}$, and potentially also the characteristics of the original vector $\mathbf{x}^*$ and the data vector $\mathbf{b}$.  Examples of this type of bound include:
\begin{itemize}
\item If $\mathbf{A}$ has full column rank   and $\hat{\mathbf{x}}$ is obtained by least-squares as $\hat{\mathbf{x}} = \mathbf{A}^\dagger\mathbf{b}$, then the spectral norm \cite{luenberger1969,golub1996,bertero1998,moon2000,heath2002,bresler2008a} provides the bound
\begin{equation}
 \|\hat{\mathbf{x}} - \mathbf{x}^*\|_2 \leq \sigma_{N}^{-1}(\mathbf{A}) \|\mathbf{n}\|_2,\label{eq:spec}
\end{equation}
where $\sigma_i(\cdot)$ denotes the $i$th largest singular value of a matrix, and $\sigma_{N}^{-1}(\mathbf{A}) = \|\mathbf{A}^\dagger\|$, where $\|\cdot\|$ denotes the spectral norm and $\mathbf{A}^\dagger$ denotes the Moore-Penrose pseudoinverse of $\mathbf{A}$. 
\item If $\mathbf{A}$ has full column rank   and $\hat{\mathbf{x}}$ is obtained by least-squares as $\hat{\mathbf{x}} =\mathbf{A}^\dagger\mathbf{b}$, then condition number analysis shows that \cite{golub1996,bertero1998,moon2000,heath2002,bresler2008a}
\begin{equation}
 \begin{split}
  &\frac{\|\hat{\mathbf{x}} - \mathbf{x}^* \|_2}{\|\mathbf{x}^*\|_2} \leq  \kappa(\mathbf{A}) \frac{\|\mathbf{n}\|_2}{\|\mathbf{b}^*\|_2},\label{eq:cond}
 \end{split}
\end{equation}
where $\mathbf{b}^*\triangleq\mathbf{b}-\mathbf{n}$ is the ideal measurement that would have been obtained in the absence of noise,\footnote{Note that the condition number bound that we have presented assumes that $\mathbf{b}^*$ belongs to the range of $\mathbf{A}$, which results in a simpler expression than would be obtained otherwise \cite{golub1996,heath2002,bresler2008a}. }  and $\kappa(\mathbf{A}) \triangleq \sigma_1(\mathbf{A})/\sigma_N(\mathbf{A})$ is the \emph{condition number} that provides a measure of the sensitivity of the inverse problem to perturbations of the data.  The best (smallest) possible condition number is $\kappa(\mathbf{A})=1$, while large values of the condition number imply that even relatively small amounts of noise can lead to potentially large relative errors.  
\item Many error bounds have been derived that apply under special structural assumptions about $\mathbf{x}^*$, such as sparsity \cite{candes2008c} or low-rank \cite{davenport2016} assumptions.  While we make no attempt to be comprehensive, a typical example is the work of Donoho \cite{donoho2006b}, which showed that if $\|\mathbf{n}\|_2 \leq \varepsilon$  and if $\mathbf{x}^*$ is sufficiently sparse, then using
$\hat{\mathbf{x}} = \arg\min_{\mathbf{x} \in \mathbb{R}^N} \|\mathbf{x}\|_1$ subject to  $\|\mathbf{A}\mathbf{x} - \mathbf{b}\|_2 \leq \varepsilon$
will result in a solution that obeys a bound with the same form as Eq.~\eqref{eq:global} for most large underdetermined matrices $\mathbf{A}$.
\end{itemize}

We refer to all of these as global bounds because they only provide insight into the aggregate error across all of the entries of the vector $\hat{\mathbf{x}}$, in contrast to an entrywise bound that would provide insight into the potential errors for each of the $N$ individual entries of the vector $\hat{\mathbf{x}}$.  

In practical applications, there are many situations where certain entries of $\mathbf{x}^*$ are easier to estimate than others and also many situations where certain entries of $\mathbf{x}^*$ are more important to estimate accurately.  In such situations, global performance bounds can be misleading, and it can be very valuable to know which entries of an obtained solution vector are more trustworthy.  For example, this situation occurs frequently in biomedical imaging applications, where the data acquisition procedure can capture more information about certain image voxels than it does for others, leading to spatially-varying ambiguity about the reconstructed image.  In such applications, this spatially-varying ambiguity has often previously been evaluated using statistical characterizations like variance maps or Cram\'{e}r-Rao bounds \cite{fessler1996a,pruessmann1999,qi2000,stayman2004,robson2008,ahn2008,haldar2018}. Use of such techniques generally requires either explicit modeling of the noise statistics or multiple repetitions of the same data collection procedure to enable empirical variance estimation, and can involve restrictive assumptions about the matrix $\mathbf{A}$ (e.g., that it has full column rank).  

In this work, we  derive novel   tight entrywise bounds that apply to every nearly data-consistent vector $\mathbf{x}$, can be used with arbitrary matrices $\mathbf{A}$ (including rank-deficient matrices), are agnostic to the estimation method used to obtain $\hat{\mathbf{x}}$, and only require knowledge about the size of $\mathbf{n}$ (via an upper bound on $\|\mathbf{n}\|_2$) without requiring an accurate statistical model or multiple repetitions of the data collection procedure.  To be precise about what we mean by ``nearly data-consistent solutions,'' we will define $\Gamma_\varepsilon(\mathbf{A},\mathbf{b})$ as the set of potential solution vectors $\mathbf{x}$ that are  consistent with the data vector $\mathbf{b}$ within a tolerance of $\varepsilon$:
\begin{equation}
\Gamma_\varepsilon(\mathbf{A},\mathbf{b}) = \left\{ \mathbf{x} \in \mathbb{R}^N \text{ s. t. } \|\mathbf{A}\mathbf{x}-\mathbf{b}\|_2 \leq \varepsilon\right\}.\label{eq:firstgamma}
\end{equation}
Our theoretical results will provide tight entrywise bounds that must be satisfied by all of the elements of $\Gamma_\varepsilon(\mathbf{A},\mathbf{b})$.  Notably, if $\|\mathbf{n}\|_2 \leq \varepsilon$ and data is acquired according to Eq.~\eqref{eq:model}, then the true value $\mathbf{x}^*$ must belong to $\Gamma_\varepsilon(\mathbf{A},\mathbf{b})$, and our entrywise bounds on  $\Gamma_\varepsilon(\mathbf{A},\mathbf{b})$ must also be bounds for $\mathbf{x}^*$.  

Our results (which will be given formally in the sequel as Thm.~\ref{thm:main} and its corollaries) will allow us to make useful statements such as:  \emph{ If $\mathbf{x} \in \Gamma_\varepsilon(\mathbf{A},\mathbf{b})$, then $x_i$, the $i$th entry of $\mathbf{x}$, must belong to a fixed interval $[L_i, U_i] \subset \mathbb{R}$, where the values of $L_i$ and $U_i$ can be computed explictly  based on the values of $\mathbf{A}$, $\mathbf{b}$, $\varepsilon$, and $i$.}  The values we obtain for $L_i$ and $U_i$ are also tight in the sense that we can easily identify specific values of $\mathbf{x} \in \Gamma_\varepsilon(\mathbf{A},\mathbf{b})$ that achieve the minimal and maximal values of $x_i$.   We are not aware of existing characterizations of this form, and we believe they enable fundamental new insights into the ambiguities associated with ill-posed inverse problems.

In addition, our results  also allow us to define an \emph{entrywise condition number} $\kappa_i(\mathbf{A})$ for the $i$th entry of $\mathbf{x}$. This entrywise condition number (given formally in Corollary~\ref{cor5}) will be used in a similar way the condition number $\kappa(\mathbf{A})$ from Eq.~\eqref{eq:cond}, but will provide bounds that are generally less pessimistic than Eq.~\eqref{eq:cond} (with $\kappa_i(\mathbf{A}) \leq \kappa(\mathbf{A}))$ and which explicitly capture the fact that some entries of $\mathbf{x}$ can be substantially better conditioned than others.   

During the the final stages of writing this paper, we became aware of mathematical literature involving \emph{componentwise condition numbers} \cite{higham1995,cucker2007}.   Componentwise theory is similar to our proposed entrywise theory in the sense that they both focus on the individual entries of the vector $\mathbf{x}$.  However, there are also important differences, including the fact that componentwise theory is frequently used to obtain a single global bound that summarizes the worst-case relative error across all of the individual entries of the vector, while we provide a finer-grained analysis by defining seperate/distinct bounds for each of the $N$ entries of $\mathbf{x}$ individually.  In addition, the literature on componentwise theory obtains different results using very different proof techniques compared to what we present in this work.

This paper is organized as follows. We start by introducing some additional notation in Sec.~\ref{sec:not}.  This is followed by a presentation of our principal theoretical results in Sec.~\ref{sec:theory}.  To illustrate the usefulness of our theory, we apply it to characterize an inverse problem from a biomedical imaging application in Sec.~\ref{sec:example}.  We conclude the paper with a discussion of several additional topics in Sec.~\ref{sec:disc}, including practical computational considerations for large problem sizes, connections between our proposed bounds and Cram\'{e}r-Rao bounds under white Gaussian noise assumptions, and interesting potential extensions to common scenarios where additional constraints are available beyond just near data-consistency (e.g., sparsity constraints, low-rank constraints, manifold constraints, etc.).

\section{Notation}\label{sec:not}
In addition to the notation that we have already introduced, the following notation is also used throughout the paper.  We use $\mathcal{R}(\mathbf{A})$ and $\mathcal{N}(\mathbf{A})$ to respectively denote the range space and nullspace of the matrix $\mathbf{A}$.  For a subspace $\mathcal{S}$, we use $\mathcal{S}^\perp$ to denote its orthogonal complement, use $\mathbf{P}_\mathcal{S}$ to denote its orthogonal projection matrix, and use $\dim(\mathcal{S})$ to denote its dimension.  

For a matrix $\mathbf{A} \in \mathbb{R}^{M\times N}$ of rank $r$, we can use the singular value decomposition (SVD) to express $\mathbf{A}$ as $ \mathbf{A} = \mathbf{U}\boldsymbol{\Sigma}\mathbf{V}^T$, where $\mathbf{U} \in \mathbb{R}^{M\times r}$ and $\mathbf{V} \in \mathbb{R}^{N \times r}$ are matrices with orthonormal columns (corresponding to the left and right singular vectors, respectively), and $\boldsymbol{\Sigma}\in \mathbb{R}^{r \times r}$ is a diagonal matrix with $i$th diagonal element equal to $\sigma_i(\mathbf{A})$.  The columns of $\mathbf{U}$ form an orthonormal basis for $\mathcal{R}(\mathbf{A})$, while the columns of $\mathbf{V}$ form an orthonormal basis for $\mathcal{R}(\mathbf{A}^T)$.  We can also write the matrix $\mathbf{A}$ using the extended SVD as
\begin{equation}
 \mathbf{A} = \begin{bmatrix}\mathbf{U} & \mathbf{U}_\perp\end{bmatrix}\begin{bmatrix}\boldsymbol{\Sigma}& \mathbf{0} \\ \mathbf{0} &\mathbf{0}\end{bmatrix} \begin{bmatrix} \mathbf{V}^T\\ \mathbf{V}_\perp^T\end{bmatrix},
\end{equation}
where $\mathbf{U}_\perp \in \mathbb{R}^{M \times (M-r)}$ and $\mathbf{V}_\perp \in \mathbb{R}^{N \times (N-r)}$ are  also matrices with orthonormal columns.  The columns of $\mathbf{U}_\perp$ form an orthonormal basis for $\mathcal{R}^\perp(\mathbf{A}) = \mathcal{N}(\mathbf{A}^T)$ and satisfy $\mathbf{U}^T\mathbf{U}_\perp=\mathbf{0}$, while the columns of $\mathbf{V}_\perp$ form an orthonormal basis for $\mathcal{R}^\perp(\mathbf{A}) = \mathcal{N}(\mathbf{A})$ and satisfy $\mathbf{V}^T\mathbf{V}_\perp=\mathbf{0}$.

We use $\mathbf{I}_T$ to denote the $T\times T$ identity matrix. We denote the Moore-Penrose pseudoinverse of the matrix $\mathbf{A}$ as $\mathbf{A}^\dagger = \mathbf{V}\boldsymbol{\Sigma}^{-1}\mathbf{U}^T$.  If $\mathbf{A}$ is square and nonsingular, then $\mathbf{A}^\dagger = \mathbf{A}^{-1}$.  If $\mathbf{A}$ has full column rank, then $\mathbf{A}^\dagger = (\mathbf{A}^T\mathbf{A})^{-1}\mathbf{A}^T$. Note that $\mathbf{P}_{\mathcal{R}(\mathbf{A})} = \mathbf{U}\mathbf{U}^T = \mathbf{A}\mathbf{A}^\dagger$, and that $\mathbf{P}_{\mathcal{R}^\perp(\mathbf{A})} = \mathbf{I}_M - \mathbf{A}\mathbf{A}^\dagger$.  

For the positive semidefinite matrix $\mathbf{A}^T\mathbf{A} = \mathbf{V}\boldsymbol{\Sigma}^2\mathbf{V}^T$, we define its square-root as $(\mathbf{A}^T\mathbf{A})^{\frac{1}{2}} \triangleq \mathbf{V}\boldsymbol{\Sigma}\mathbf{V}^T$.  

We will use $\mathbf{e}_i \in \mathbb{R}^N$ to denote the vector whose $i$th entry is equal to one, with all the other entries equal to zero (i.e., $\mathbf{e}_i$ is the $i$th column of $\mathbf{I}_N$).

\section{Main Results}\label{sec:theory}
Our main theoretical results are given by the following theorem and its corollaries.

\begin{theorem}
Let $\Gamma_\varepsilon(\mathbf{A},\mathbf{b})$ be defined as in Eq.~\eqref{eq:firstgamma}, and let 
\begin{equation*}
 \Phi_\varepsilon^i(\mathbf{A},\mathbf{b}) = \left\{\begin{array}{l}x_i \in \mathbb{R}  \;\;\;\text{s. t. }  x_i = \mathbf{e}_i^T\mathbf{x} \\\;\;\;\;\;\;\;\;\; \;\;\;\;\;\;\;\;\;\text{ for some } \mathbf{x} \in \Gamma_\varepsilon(\mathbf{A},\mathbf{b})\end{array}\right\}                                                                  \end{equation*}
 denote the set of all possible values for $x_i$ (the $i$th entry of $\mathbf{x}$) across the set of all nearly data-consistent vectors $\mathbf{x} \in \Gamma_\varepsilon(\mathbf{A},\mathbf{b})$.
  If $\mathbf{e}_i$ is orthogonal to  $\mathcal{N}(\mathbf{A})$, then $\Phi_\varepsilon^i(\mathbf{A},\mathbf{b})$ is the finite interval $\Phi_\varepsilon^i(\mathbf{A},\mathbf{b}) = [L_i,U_i] \subset \mathbb{R}$,  with
\begin{equation*}
\begin{split}
  L_i &= \min_{\mathbf{x} \in \mathbb{R}^N} \mathbf{e}_i^T\mathbf{x}  \,\,\,\,\, \text{ s. t. } \,\,\,\,\, \|\mathbf{A}\mathbf{x}-\mathbf{b}\|_2 \leq \varepsilon\\  
  &= \mathbf{e}_i^T\mathbf{A}^\dagger\mathbf{b}- \left\| \left(\mathbf{A}^\dagger\right)^T\mathbf{e}_i\right\|_2 \sqrt{\varepsilon^2-\|\mathbf{P}_{\mathcal{R}^\perp(\mathbf{A})}\mathbf{b}\|_2^2},  
  \end{split}
\end{equation*}
\begin{equation*}
\begin{split}
  U_i &= \max_{\mathbf{x} \in \mathbb{R}^N} \mathbf{e}_i^T\mathbf{x}  \,\,\,\,\, \text{ s. t. } \,\,\,\,\, \|\mathbf{A}\mathbf{x}-\mathbf{b}\|_2 \leq \varepsilon\\  
  &= \mathbf{e}_i^T\mathbf{A}^\dagger\mathbf{b}+ \left\| \left(\mathbf{A}^\dagger\right)^T\mathbf{e}_i\right\|_2 \sqrt{\varepsilon^2-\|\mathbf{P}_{\mathcal{R}^\perp(\mathbf{A})}\mathbf{b}\|_2^2},
  \end{split}
\end{equation*}
and midpoint $\frac{1}{2}(L_i+U_i) = \mathbf{e}_i^T\mathbf{A}^\dagger\mathbf{b}$.
Conversely, if $\mathbf{e}_i$ is not orthogonal to $\mathcal{N}(\mathbf{A})$, then $\Phi_\varepsilon^i(\mathbf{A},\mathbf{b})$ is unbounded with $\Phi_\varepsilon^i(\mathbf{A},\mathbf{b}) = \mathbb{R}$.\label{thm:main}
\end{theorem}

This theorem is valid for arbitrary $\mathbf{A}$, $\mathbf{b}$, and $\varepsilon$, and shows that, depending on how the nullspace  of the matrix $\mathbf{A}$ interacts with the vector $\mathbf{e}_i$, the value of $x_i$ is either restricted to a known interval of length $2\left\| \left(\mathbf{A}^\dagger\right)^T\mathbf{e}_i\right\|_2\sqrt{\varepsilon^2-\|\mathbf{P}_{\mathcal{R}^\perp(\mathbf{A})}\mathbf{b}\|_2^2}$ or is entirely unconstrained and can be an arbitrary real number.

While Thm.~\ref{thm:main} applies to the general case, simplifications occur under  assumptions about the rank  of $\mathbf{A}$. Specifically:
\begin{itemize}
 \item If $\mathbf{A}$ has full column rank (i.e., $\mathrm{rank}(\mathbf{A}) = N$), then $\mathbf{e}_i$ is always orthogonal to $\mathcal{N}(\mathbf{A})$, and $\Phi_\varepsilon^i(\mathbf{A},\mathbf{b})$ is always a finite interval.  
 \item If $\mathbf{A}$ has full row rank (i.e., $\mathrm{rank}(\mathbf{A}) = M$), then $\|\mathbf{P}_{\mathcal{R}^\perp(\mathbf{A})}\mathbf{b}\|_2$ is always zero.  This implies that $\Phi_\varepsilon^i(\mathbf{A},\mathbf{b})$ is a finite interval of length $2\varepsilon\left\|  \left(\mathbf{A}^\dagger\right)^T\mathbf{e}_i\right\|_2$ when $\mathbf{e}_i$ is orthogonal to $\mathcal{N}(\mathbf{A})$, and is unbounded with $\Phi_\varepsilon^i(\mathbf{A},\mathbf{b}) = \mathbb{R}$ when $\mathbf{e}_i$ is not orthogonal to $\mathcal{N}(\mathbf{A})$.
 \item If $\mathbf{A}$ is square and non-singular (i.e., $\mathrm{rank}(\mathbf{A}) = N = M$), then $\Phi_\varepsilon^i(\mathbf{A},\mathbf{b})$ is always the finite interval of length  $2\varepsilon\left\| (\mathbf{A}^{-1})^T\mathbf{e}_i\right\|_2$ with midpoint $\mathbf{e}_i^T\mathbf{A}^{-1}\mathbf{b}$.  
\end{itemize}

It should also be noted that Thm.~\ref{thm:main} will still be valid if the vector $\mathbf{e}_i$ is substituted everywhere with an arbitrary nonzero vector $\mathbf{w} \in \mathbb{R}^N$.  This allows the use of these kinds of bounds in the case where we are potentially interested in the range of values for a weighted linear combination of the entries of $\mathbf{x}$ in the form $\mathbf{w}^T\mathbf{x}$.  For example, in practical imaging applications, if voxel $x_i$ appears to have a larger value than a neighboring voxel $x_j$ in the reconstructed image, it may be worthwhile to know tight upper and lower bounds on the value of $x_i - x_j$ across all  $\mathbf{x} \in  \Gamma_\varepsilon(\mathbf{A},\mathbf{b})$, which  would allow direct insight into whether voxel $x_i$ is always larger than voxel $x_j$ within the class of nearly data-consistent images.   This could easily be achieved using the weight vector $\mathbf{w} = \mathbf{e}_i - \mathbf{e}_j$. 

Before giving the proof of Thm.~\ref{thm:main}, we will first remark that minor manipulations of Thm.~\ref{thm:main} also lead to the following corollaries.  (As with Thm.~\ref{thm:main}, these corollaries are also valid if the vector $\mathbf{e}_i$ is replaced everywhere with an arbitrary nonzero vector $\mathbf{w}\in\mathbb{R}^N$).
\begin{corollary}
 Assume that noisy data $\mathbf{b}$ is measured according to Eq.~\eqref{eq:model}, and that the noise is known to obey $\|\mathbf{n}\|_2 \leq \varepsilon$. Let $x_i^* \triangleq \mathbf{e}_i^T\mathbf{x}^*$ denote the $i$th entry of the original true vector $\mathbf{x}^*$.  Then $x_i^* \in \Phi_\varepsilon^i(\mathbf{A},\mathbf{b})$, with $ \Phi_\varepsilon^i(\mathbf{A},\mathbf{b})$  defined as in Thm.~\ref{thm:main}.
\end{corollary}
\begin{corollary}
Assume that noisy data $\mathbf{b}$ is measured according to Eq.~\eqref{eq:model}, and that the noise is known to obey $\|\mathbf{n}\|_2 \leq \varepsilon$.  Also assume that an estimation procedure is used to obtain an estimate $\hat{\mathbf{x}}$ of $\mathbf{x}^*$, and that this estimate is nearly data-consistent such that $\hat{\mathbf{x}}\in \Gamma_\varepsilon(\mathbf{A},\mathbf{b})$. Let $\hat{x}_i \triangleq \mathbf{e}_i^T\hat{\mathbf{x}}$ and $x_i^* \triangleq \mathbf{e}_i^T\mathbf{x}^*$ denote the $i$th entries of the estimate and the original true vector, respectively. Then if $\mathbf{e}_i$ is orthogonal to $\mathcal{N}(\mathbf{A})$, we must have
\begin{equation*}
\begin{split}
 \left|\hat{x}_i - x_i^*\right| & \leq \left\| \left(\mathbf{A}^\dagger\right)^T\mathbf{e}_i\right\|_2\sqrt{\varepsilon^2-\|\mathbf{P}_{\mathcal{R}^\perp(\mathbf{A})}\mathbf{b}\|_2^2}\\ &\;\;\;\;\;\;\;\; +\left|\hat{x}_i - \mathbf{e}_i^T\mathbf{A}^\dagger\mathbf{b}\right|\\
 &\leq 2 \left\| \left(\mathbf{A}^\dagger\right)^T\mathbf{e}_i\right\|_2\sqrt{\varepsilon^2-\|\mathbf{P}_{\mathcal{R}^\perp(\mathbf{A})}\mathbf{b}\|_2^2}.
 \end{split}
\end{equation*}
\end{corollary}
\begin{corollary}\label{cor3}
Assume that noisy data $\mathbf{b}$ is measured according to Eq.~\eqref{eq:model},  that $\mathbf{A}$ has full column rank, and that $\hat{\mathbf{x}}$ is obtained using $\hat{\mathbf{x}} = \mathbf{A}^\dagger\mathbf{b}$.  Then 
\begin{equation*}
\begin{split}
 \left|\hat{x}_i - x_i^*\right| & \leq \left\| \left(\mathbf{A}^\dagger\right)^T\mathbf{e}_i\right\|_2\sqrt{\|\mathbf{n}\|_2^2-\|\mathbf{P}_{\mathcal{R}^\perp(\mathbf{A})}\mathbf{b}\|_2^2}\\
 &\leq \left\| \left(\mathbf{A}^\dagger\right)^T\mathbf{e}_i\right\|_2  \|\mathbf{n}\|_2.
 \end{split}
\end{equation*}
\end{corollary}

Note that Cor.~\ref{cor3} can be viewed an entrywise analogue of the spectral norm bound from Eq.~\eqref{eq:spec}. Moreover, Cor.~\ref{cor3} provides a bound that is generally better and never worse than Eq.~\eqref{eq:spec}, since we always have that $\left\| \left(\mathbf{A}^{\dagger}\right)^T\mathbf{e}_i\right\|_2 \leq \sigma_N^{-1}(\mathbf{A})$.

\begin{corollary}\label{cor5}
Assume that noisy data $\mathbf{b}$ is measured according to Eq.~\eqref{eq:model},  that $\mathbf{A}$ has full column rank, and that $\hat{\mathbf{x}}$ is obtained using $\hat{\mathbf{x}} = \mathbf{A}^\dagger\mathbf{b}$.  Then 
\begin{equation*}
\begin{split}
 \frac{\left|\hat{x}_i - x_i^*\right|}{\|\mathbf{x}^*\|_2} & \leq \kappa_i(\mathbf{A})\frac{\sqrt{\|\mathbf{n}\|_2^2-\|\mathbf{P}_{\mathcal{R}^\perp(\mathbf{A})}\mathbf{b}\|_2^2}}{\|\mathbf{b}^*\|_2}\\
 & \leq \kappa_i(\mathbf{A}) \frac{\|\mathbf{n}\|_2}{\|\mathbf{b}^*\|_2},
 \end{split}
\end{equation*}
where we define $\kappa_i(\mathbf{A}) \triangleq \left\| \left(\mathbf{A}^\dagger\right)^T\mathbf{e}_i\right\|_2\sigma_1(\mathbf{A})$ as the \emph{entrywise condition number} for the $i$th entry of $\mathbf{x}$.
\end{corollary}

Note that Cor.~\ref{cor5} can be viewed as an entrywise analogue of the condition number bound from Eq.~\eqref{eq:cond}.  In addition, Cor.~\ref{cor5} provides a bound that is generally better and never worse than Eq.~\eqref{eq:cond}, since we always have that $\kappa_i(\mathbf{A}) \leq \kappa(\mathbf{A})$ (which can be established from the fact that $\left\| \left(\mathbf{A}^{\dagger}\right)^T\mathbf{e}_i\right\|_2 \leq \sigma_N^{-1}(\mathbf{A})$).

The proof of Thm.~\ref{thm:main} is given below.

\subsection*{Proof of Thm.~\ref{thm:main}}
We first decompose $\mathbf{b}$ as $\mathbf{b} = \mathbf{P}_{\mathcal{R}(\mathbf{A})} \mathbf{b} + \mathbf{P}_{\mathcal{R}^\perp(\mathbf{A})} \mathbf{b} = \mathbf{A}\mathbf{A}^\dagger \mathbf{b} + \mathbf{P}_{\mathcal{R}^\perp(\mathbf{A})} \mathbf{b}$.  Due to orthogonality between $\mathcal{R}(\mathbf{A})$ and $\mathcal{R}^\perp(\mathbf{A})$ (invoking the Pythagorean theorem for $\mathbb{R}^N$), we can rewrite $\Gamma_\varepsilon(\mathbf{A},\mathbf{b})$ as 
\begin{equation}
\begin{split}
\Gamma_\varepsilon(\mathbf{A},\mathbf{b}) &= \left\{\begin{array}{l} \mathbf{x} \in \mathbb{R}^N  \\ \; \text{s. t. } \|\mathbf{A}\mathbf{x} - \mathbf{A}\mathbf{A}^\dagger \mathbf{b}\|_2^2 + \|\mathbf{P}_{\mathcal{R}^\perp(\mathbf{A})} \mathbf{b}\|_2^2 \leq \varepsilon^2 \end{array} \right\}\\
&= \left\{\mathbf{x} \in \mathbb{R}^N \text{ s. t. } \|\mathbf{A}\mathbf{x} - \mathbf{A}\mathbf{z}\|_2^2 \leq \lambda^2  \right\},
\end{split}
\end{equation}
with $\lambda^2 \triangleq \varepsilon^2  - \|\mathbf{P}_{\mathcal{R}^\perp(\mathbf{A})} \mathbf{b}\|_2^2$ and $\mathbf{z} \triangleq \mathbf{A}^\dagger\mathbf{b}$. 
Note that $\|\mathbf{P}_{\mathcal{R}^\perp(\mathbf{A})} \mathbf{b}\|_2$ must be smaller than $\varepsilon$ for $\Gamma_\varepsilon(\mathbf{A},\mathbf{b})$ to be a nonempty set.

Using the definition of the $\ell_2$-norm, we can simplify the previous expression to
\begin{equation}
\begin{split}
\Gamma_\varepsilon(\mathbf{A},\mathbf{b}) = \left\{\begin{array}{l} \mathbf{x} \in \mathbb{R}^N \\ \;\;\; \text{s. t. } \left\|\left(\frac{\mathbf{A}^T\mathbf{A}}{\lambda^2}\right)^{\frac{1}{2}}(\mathbf{x} - \mathbf{z})\right\|_2^2 \leq 1\end{array}\right\}.
\end{split}
\end{equation}
This set has the form of an ellipsoid \cite{boyd2004}, and it is well-known that if $\mathbf{A}^T\mathbf{A}$ is nonsingular, then  the volume of this ellipsoid is \cite{friendly2011}
\begin{equation}
\frac{\pi^{\frac{N}{2}} \lambda^N}{\Gamma(\frac{N}{2} +1)}\sqrt{\left|(\mathbf{A}^T\mathbf{A})^{-1}\right|},
\end{equation}
where $\Gamma(\cdot)$ denotes the gamma function (and should not be confused with the set $\Gamma_\varepsilon(\mathbf{A},\mathbf{b})$).  Clearly, if the volume of this ellipsoid is large, then there will be many vectors that are nearly data-consistent, and the solution to the inverse problem will have more potential ambiguity.  If $\mathbf{A}^T\mathbf{A}$ were singular, then the ellipsoid is degenerate and must extend infinitely along directions aligned with the nullspace of $\mathbf{A}$, causing $\Gamma_\varepsilon(\mathbf{A},\mathbf{b})$ to have infinite volume.  However, these statements only provide insight into the global ambiguity of the solution, and our  goal in this work is to characterize the ambiguities associated with the individual entries of the elements of $\Gamma_\varepsilon(\mathbf{A},\mathbf{b})$, which may be much better for some entries than might be interpreted based on a global characterization.

Invoking the SVD  $\mathbf{A} = \mathbf{U}\boldsymbol{\Sigma}\mathbf{V}^T$, we have
\begin{equation}
\begin{split}
\Gamma_\varepsilon(\mathbf{A},\mathbf{b}) 
&=  \left\{\mathbf{x} \in \mathbb{R}^N \text{ s. t. } \left\|\frac{\boldsymbol{\Sigma}\mathbf{V}^T}{\lambda}(\mathbf{x} - \mathbf{z})\right\|_2^2 \leq  1\right\}.
\end{split}
\end{equation}
Furthermore, because of the orthonormality and mutual orthogonality properties of $\mathbf{V}$ and $\mathbf{V}_\perp$, it is possible to uniquely represent an arbitrary vector $\mathbf{x} \in \mathbb{R}^N$ as $\mathbf{x} = \mathbf{V}\mathbf{p} + \mathbf{V}_\perp \mathbf{q}$, where $\mathbf{p} \in \mathbb{R}^r$ with $\mathbf{p} = \mathbf{V}^T\mathbf{x}$, and $\mathbf{q} \in \mathbb{R}^{N-r}$ with $\mathbf{q} = \mathbf{V}_\perp^T\mathbf{x}$.  We can then write
\begin{equation}
\begin{split}
\Gamma_\varepsilon(\mathbf{A},\mathbf{b}) 
&=  \left\{ \begin{array}{l} \mathbf{x} = \mathbf{V}\mathbf{p} + \mathbf{V}_\perp\mathbf{q}  \\\;\;\;\text{s. t. }
\mathbf{p} \in \mathbb{R}^r, \mathbf{q} \in \mathbb{R}^{N-r}, \\\;\;\;\;\;\;\;\;\;\; \text{and } \left\|\frac{\boldsymbol{\Sigma}}{\lambda}(\mathbf{p} - \mathbf{V}^T\mathbf{z})\right\|_2^2 \leq  1 \end{array} \right\}.
\end{split}
\end{equation}
A simple change of variables then results in
\begin{equation}
\begin{split}
\Gamma_\varepsilon(\mathbf{A},\mathbf{b}) 
&=  \left\{ \begin{array}{l} \mathbf{x} = \lambda\mathbf{V}\boldsymbol{\Sigma}^{-1}\mathbf{p}+ \mathbf{V}_\perp\mathbf{q} + \mathbf{z}   \\ \;\;\;   \text{s. t. } \mathbf{p} \in \mathbb{R}^r, \mathbf{q} \in \mathbb{R}^{N-r}, \\ \;\;\;\;\;\;\;\;\;\; \text{and } \left\|\mathbf{p}\right\|_2^2 \leq  1 \end{array} \right\}.
\end{split}\label{eq:finalgamma}
\end{equation}
Notably, the component $\mathbf{q}$ (which corresponds to the part of $\mathbf{x}$ that is in the nullspace of $\mathbf{A}$) is completely unconstrained, and can be arbitrarily large. This results in a degenerate ellipsoid as described previously.  

Given this geometric characterization of the set of nearly data-consistent solutions, we can now consider the behavior of the individual entries of $\mathbf{x}$. 
Substituting the results of Eq.~\eqref{eq:finalgamma} into the definition of $\Phi_\varepsilon^i(\mathbf{A},\mathbf{b})$ as given in Thm~\ref{thm:main}, we obtain 
\begin{equation}
 \Phi_\varepsilon^i(\mathbf{A},\mathbf{b}) \hspace{-0.3em}= \hspace{-0.3em}\left\{\begin{array}{l}\hspace{-0.3em}x_i = \lambda\mathbf{e}_i^T\mathbf{V}\boldsymbol{\Sigma}^{-1}\mathbf{p} + \mathbf{e}_i^T\mathbf{V}_\perp\mathbf{q} + \mathbf{e}_i^T\mathbf{z}  \\ \;\;\;   \text{s. t. } \mathbf{p} \in \mathbb{R}^r, \mathbf{q} \in \mathbb{R}^{N-r}, \\ \;\;\;\;\;\;\;\;\;\; \text{and } \left\|\mathbf{p}\right\|_2^2 \leq  1 \end{array} \right\}.
\end{equation}

At this point, there are two cases to consider.  
\subsubsection{Case 1} First, consider the case where $\mathbf{e}_i$ is not orthogonal to $\mathcal{N}(\mathbf{A})$, such that $\mathbf{e}_i^T \mathbf{V}_\perp \neq \mathbf{0}$. Then, because $\mathbf{q}$ is unconstrained, we must have 
\begin{equation}
 \Phi_\varepsilon^i(\mathbf{A},\mathbf{b}) = \mathbb{R},
\end{equation}
indicating that the $i$th entry of $\mathbf{x}$ is unbounded and could take on arbitrary values.  In particular, for an arbitrary $\alpha \in \mathbb{R}$, we can achieve $\alpha = x_i = \mathbf{e}_i^T\mathbf{x}$ for $\mathbf{x} \in \Gamma_\varepsilon(\mathbf{A},\mathbf{b})$ by choosing $\mathbf{x} = \mathbf{V}_\perp\mathbf{q}+\mathbf{z}$ with $\mathbf{q} = (\alpha-\mathbf{e}_i^T\mathbf{z}) \mathbf{V}_\perp^T\mathbf{e}_i/\left\|\mathbf{V}_\perp^T\mathbf{e}_i\right\|_2^2$.

\subsubsection{Case 2}In the second case, assume that $\mathbf{e}_i$  is orthogonal to $\mathcal{N}(\mathbf{A})$ such that $\mathbf{e}_i^T \mathbf{V}_\perp = \mathbf{0}$.  In this case, we have the simplification
\begin{equation}
 \Phi_\varepsilon^i(\mathbf{A},\mathbf{b}) = \left\{\begin{array}{l}x_i = \lambda\mathbf{e}_i^T\mathbf{V}\boldsymbol{\Sigma}^{-1}\mathbf{p} + \mathbf{e}_i^T\mathbf{z}  \\ \;\;\;   \text{s. t. } \mathbf{p} \in \mathbb{R}^r \text{ and } \left\|\mathbf{p}\right\|_2^2 \leq  1 \end{array} \right\}.
\end{equation}
The maximum and minimum possible values of $\lambda\mathbf{e}_i^T\mathbf{V}\boldsymbol{\Sigma}^{-1}\mathbf{p}$ under the constraint that $\left\|\mathbf{p}\right\|_2^2 \leq  1$ are respectively given by $\pm \sigma_1\left(\lambda\mathbf{e}_i^T\mathbf{V}\boldsymbol{\Sigma}^{-1}\right)= \pm\lambda\|\boldsymbol{\Sigma}^{-1}\mathbf{V}^T\mathbf{e}_i\|_2=\pm\lambda\left\|\left(\mathbf{A}^\dagger\right)^T\mathbf{e}_i\right\|_2$. Note also that the extremal values of $\lambda\mathbf{e}_i^T\mathbf{V}\boldsymbol{\Sigma}^{-1}\mathbf{p}$ are achieved  by taking  $\mathbf{p} = \pm \boldsymbol{\Sigma}^{-1} \mathbf{V}^T\mathbf{e}_i/\left\| \boldsymbol{\Sigma}^{-1} \mathbf{V}^T\mathbf{e}_i\right\|_2$, which satisfies $\|\mathbf{p}\|_2=1$, and that simple rescaling of this choice of $\mathbf{p}$ by a real-scalar with modulus less than one will allow us to achieve any real value in between the upper and lower bounds. We therefore find that $ \Phi_\varepsilon^i(\mathbf{A},\mathbf{b})$ is simply the finite interval $\Phi_\varepsilon^i(\mathbf{A},\mathbf{b}) =  [L_i,U_i] \subset \mathbb{R}$, with $L_i$ and $U_i$ as given in the statement of the theorem.  

This completes the proof of Thm.~\ref{thm:main}.  $\square$

While most of the corollaries are trivial manipulations of Thm.~\ref{thm:main}, we will note that the derivation of  Cor.~\ref{cor5} is based on the combination of Cor.~\ref{cor3} with the fact that $\|\mathbf{b}^*\|_2 = \|\mathbf{A}\mathbf{x}^*\|_2 \leq \sigma_1(\mathbf{A})\|\mathbf{x}^*\|_2.$

\section{Illustrative Example}\label{sec:example}
We illustrate the utility of our new theoretical results by applying them to an example application: image reconstruction in multi-channel magnetic resonance imaging (MRI).  Specifically, under the sensitivity encoding model of multi-channel MRI \cite{pruessmann1999,pruessmann2001}, we assume that we are interested in reconstructing a discrete image defined on an $N$-voxel grid, where the value of the image at each voxel is complex-valued.  We use $\mathbf{x} \in \mathbb{C}^{N}$ to represent the unknown voxel values.  Data measurements are then obtained from this image from a series of $L$ channels, where the data $\mathbf{b}_\ell$ from the $\ell$th channel is represented as
\begin{equation}
 \mathbf{b}_\ell = \mathbf{F}\mathbf{S}_\ell \mathbf{x} + \mathbf{n}_\ell, \text{ for } \ell=1,\ldots,L,
\end{equation}
where $\mathbf{n}_\ell$ represents the noise for the $\ell$th channel, $\mathbf{S}_\ell\in \mathbb{C}^{N\times N}$ is a diagonal matrix whose diagonal entries are equal to the corresponding values of the spatially-varying sensitivity profile for the $\ell$th channel, and $\mathbf{F} \in \mathbb{C}^{P \times N}$ represents an operator that performs  Fourier transformation followed by some form of sampling.  Typically, the sampling rate is chosen to be below the Nyquist rate to allow accelerated data acquisition, resulting in $P < N$.    This allows us to represent the inverse problem as
\begin{equation}
 \mathbf{b} = \mathbf{A}\mathbf{x} + \mathbf{n},
\end{equation}
where $\mathbf{b} \in \mathbb{C}^{M}$ is the vertical concatenation of the $\mathbf{b}_\ell$ vectors, $\mathbf{n} \in \mathbb{C}^{M}$ is the vertical concatenation of the $\mathbf{n}_\ell$ vectors, and  $\mathbf{A} \in \mathbb{C}^{M \times N}$ is the vertical concatenation of the $\mathbf{F}\mathbf{S}_\ell$ matrices, with $M=LP$. 

\begin{figure}[t]
\begin{minipage}{\columnwidth}
\centering
 {\color{white}\,}\hfill \includegraphics[width=1.5in]{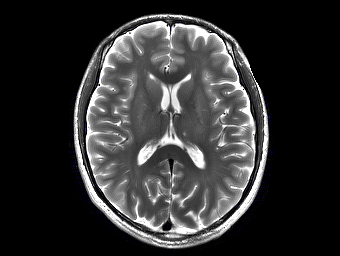} \hfill  \includegraphics[width=1.5in]{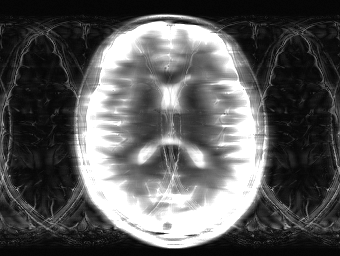} \hfill {\color{white}\,}\\[0.2em]
 {\color{white}\,}\hfill \includegraphics[width=1.5in]{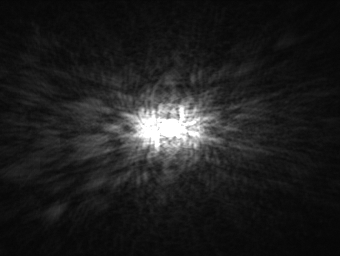} \hfill  \includegraphics[width=1.5in]{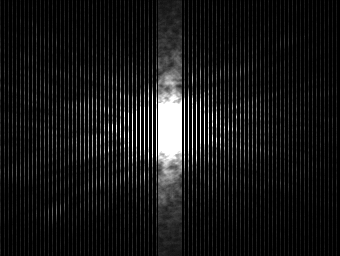} \hfill {\color{white}\,}\\ {\color{white}\,}\hspace{-0.05in}
 \hfill \small(a) Fully-Sampled \hfill (b) Subsampled + Zero-Filling \hfill \hspace{-0.35in} {\color{white}\,}
 \end{minipage}
 \caption{The (a) original fully-sampled 32-channel MRI brain data and (b) the retrospectively subsampled version of the data used for the illustrative MRI reconstruction example.  The top row shows coil-combined magnitude images, with zero-filling of the missing Fourier data samples in the subsampled case. The bottom row shows the corresponding coil-combined Fourier magnitudes.  The Fourier sampling pattern captures every fourth phase-encoding line, while also measuring the central (low-frequency) 24 phase encoding lines.  } 
 \label{fig:zerofill}
\end{figure} 

An important observation is that the matrices and vectors in our example inverse problem are all complex-valued, while our theoretical results are all given assuming the real-valued case.  We chose to present our theory for the real-valued case because there exists a standard ordering for real-numbers that enables a simple discussion of lower- and upper-bounds, while the complex-valued case would be much more complicated.  However, in order to apply our theory to this scenario, we need to transform the complex-valued problem statement into an equivalent real-valued problem statement.  This is easily done by separating each complex-value into its real- and imaginary-components, and then working with an equivalent purely-real expression of the inverse problem (see, e.g., \cite{bydder2005,kim2022}). We have used this real-valued transformation approach in the results that follow.  For simplicity and to avoid introducing new notation, we will simply refer to these now real-valued variables using the same notation we used for the complex-valued case (i.e., $\mathbf{x}$, $\mathbf{A}$, $\mathbf{b}$, etc.). From this point forward, all variables should now be interpreted as real-valued.

In our illustrative example application, we use real fully-sampled 32-channel MRI brain data acquired at our institution, which we retrospectively subsample to emulate an accelerated data acquisition procedure as illustrated in Fig.~\ref{fig:zerofill}, with $P \approx N/3.3$.  Note that even though the matrix $\mathbf{F}$ is underdetermined because $P < N$, the overall system $\mathbf{A}$ is still overdetermined because $LP > N$. We have determined numerically that the matrix $\mathbf{A}$ has full column rank.

As can be seen in Fig.~\ref{fig:zerofill}, our subsampling scheme is designed such that undersampling only occurs along one dimension of the 2D Fourier domain (i.e., the \emph{phase encoding dimension} in MRI terminology), while the remaining dimension (i.e., the \emph{readout dimension} in MRI terminology) is fully sampled at the Nyquist rate.  This structure implies that one can apply a unitary inverse Fourier transform operator along the fully-sampled readout dimension as a preprocessing step to completely decouple the reconstruction of each image row \cite{kyriakos2000}.  Reconstructing each image row independently enables substantial reductions in computational complexity, and the results we present later in this section are all based on this row-decoupled approach. 

In our illustrative example, we have used ESPIRiT \cite{uecker2014} with circular neighborhoods \cite{lobos2022} to estimate the sensitivity profiles needed to form the $\mathbf{S}_\ell$ matrices from the fully-sampled central Fourier data.  In addition, we incorporate a phase estimate directly into the sensitivity maps so that the reconstructed image will be approximately real-valued \cite{bydder2005,willig2005,lew2007,haldar2012,blaimer2016}. Our formulation also uses prior knowledge of the support of the image and does not estimate voxel locations that are already known to be zero, which is achieved by removing the corresponding columns from the $\mathbf{A}$ matrix \cite{pruessmann1999}.  We have also used direct measurements of the multi-channel thermal noise to prewhiten the channels, which allows all thermal noise samples to be treated as independent and identically distributed zero-mean circularly-symmetric complex Gaussian noise \cite{pruessmann2001}.

Our illustration will start  by calculating the theoretical lower and upper bounds from Thm.~\ref{thm:main} for this inverse problem.  However, before this can be done, it is important to select an appropriate value of $\varepsilon$.  Ideally, the noise perturbation $\mathbf{n}$ in MRI would be purely due to thermal (white Gaussian) noise, such that the distribution of $\|\mathbf{n}\|_2$ would follow the chi distribution, which has a well-characterized cumulative distribution function (CDF).  If this were the case, then a reasonable approach for selecting  $\varepsilon$ might be to choose $\varepsilon$ based on the CDF, e.g., choose $\varepsilon$ so that the probability of observing $\|\mathbf{n}\|_2 > \varepsilon$ is less than $1\%$.  Unfortunately, this approach for determining $\varepsilon$ turns out to be inappropriate for this data, since we observe empircally that we would have $\|\mathbf{P}_{\mathcal{R}^\perp(\mathbf{A})}\mathbf{b}\|_2 > \varepsilon$ under this approach to choosing $\varepsilon$.  The reason this approach fails is that we are working with real data, and there are many sources of error in the data beyond just the thermal noise, including minor motion of the human subject during the scan, flow of the subject's blood and cerebrospinal fluid during the scan, and spin relaxation during the acquisition.  All of these contribute to $\mathbf{n}$, causing it to be much larger than would be predicted from the statistics of thermal noise.  This example also helps to underscore the point from the introduction that it can be important to have theoretical bounds that are not heavily dependent on statistical noise modeling assumptions, since these assumptions can fail for practical real-world data.

To overcome this practical issue, we have used a heuristic (and likely suboptimal) approach for choosing $\varepsilon$, that is nevertheless good enough for our illustration despite its imperfections.  Specifically, we observe that, based on our inverse problem model from Eq.~\eqref{eq:model}, the vector $\mathbf{P}_{\mathcal{R}^\perp(\mathbf{A})}\mathbf{b}$ should not contain any of the true signal, and consists entirely of the projection of the  noise $\mathbf{n}$ onto the subspace $\mathcal{R}^\perp(\mathbf{A})$.  If we make the assumption that the noise energy $\mathbf{n}$ distributes relatively evenly across all subspaces of $\mathbb{R}^M$ and that $\dim(\mathbb{R}^M)=M$ and $\dim(\mathcal{R}^\perp(\mathbf{A})) = M-N$ are large enough that concentration of measure principles can be applied, then we might reasonably expect that 
\begin{equation}
\begin{split}
&\frac{\|\mathbf{n}\|_2^2}{\dim(\mathbb{R}^M)} \approx                               \frac{\|\mathbf{P}_{\mathcal{R}^\perp(\mathbf{A})}\mathbf{b}\|_2^2}{\dim(\mathcal{R}^\perp(\mathbf{A}))}                                                                                                                                                                                                                                                                                                                                                                                                                                                                                                                                                                                                                                                                                                                                                                                                                                                                                                                                              \\
\implies & \|\mathbf{n}\|_2 \approx \sqrt{\frac{M}{M-N}}\|\mathbf{P}_{\mathcal{R}^\perp(\mathbf{A})}\mathbf{b}\|_2.
\end{split}
\end{equation}
In what follows, we therefore set $\varepsilon = \sqrt{\frac{M}{M-N}}\|\mathbf{P}_{\mathcal{R}^\perp(\mathbf{A})}\mathbf{b}\|_2$.  Before moving on, we would like to reemphasize that this choice is heuristic and not rigorously justified, and many other ways of choosing $\varepsilon$ could have been used instead.  Different choices of $\varepsilon$ would result in a very simple and predictable change in the lower- and upper-bounds $L_i$ and $U_i$, and would have no effect on other properties that may be of interest such as the entrywise spectral norm-type bound $\|\left(\mathbf{A}^{\dagger}\right)^T\mathbf{e}_i\|_2$ from Cor.~\ref{cor3}  or the entrywise condition number $\kappa_i(\mathbf{A})$ from Cor.~\ref{cor5}. 

\begin{figure*}[t]
\begin{minipage}{\textwidth}
 \includegraphics[width=\textwidth]{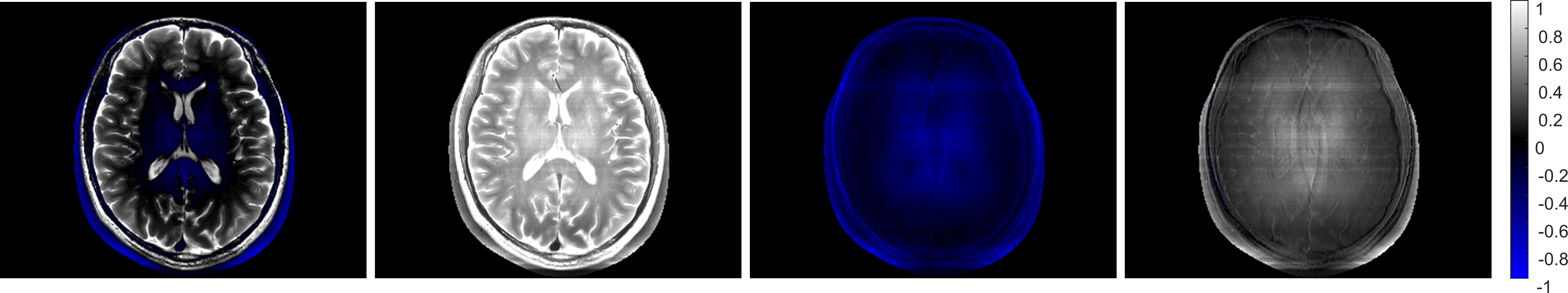}\\[-0.5em]
 \small {\color{white}\,} \hspace{0.3in} (a) $L_i$ (real part) \hspace{0.8in} (b) $U_i$ (real part) \hspace{0.6in} (c) $L_i$ (imaginary part) \hspace{0.4in} (d) $U_i$ (imaginary part)
\end{minipage}
 \caption{Illustration of spatial maps of the entrywise lowerbounds $L_i$ and upperbounds $U_i$ for the (a,b) real  and (c,d) imaginary  parts of nearly data-consistent images $\mathbf{x}\in\Gamma_\varepsilon(\mathbf{A},\mathbf{b})$, as obtained by applying Thm.~\ref{thm:main} to the multi-channel MRI reconstruction scenario.}
 \label{fig:bounds}
\end{figure*}

Figure~\ref{fig:bounds} depicts the entrywise lower- and upper-bounds we obtain on the real- and imaginary-parts of the image based on our choice of $\varepsilon$.  In this example, the lower-bounds and upper-bounds are relatively far separated from one another for most voxels of the images, suggesting that near data-consistency constraints themselves may be insufficient to guarantee accurate estimation of the image voxel amplitudes.  Of course, it should also be kept in mind that these are worst-case bounds that are independent of any specific image estimation method, and that  the practical performance of a specific estimator in the presence of typical (i.e., not worst-case) $\mathbf{n}$ may be better.  On the other hand, the fact that there can be this level of variation within the set of near data-consistent images may also encourage an experimenter to consider the use of noise-mitigation strategies so that $\varepsilon$ can be reduced, and/or the redesign of the data acquisition matrix $\mathbf{A}$ so that $\|(\mathbf{A}^\dagger)^T\mathbf{e}_i\|_2$ can be made smaller.

\begin{figure}[t]
\centering
\includegraphics[width=0.9\columnwidth]{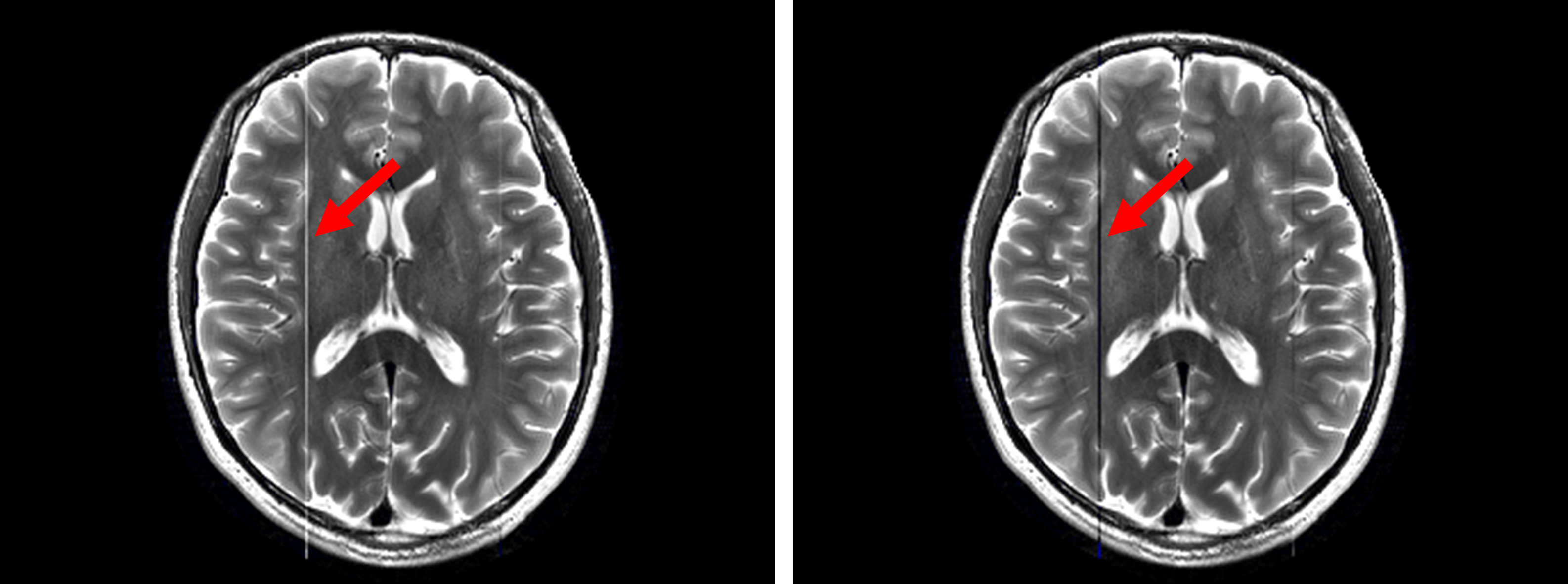}\\
{\color{white}\,}\hspace{-0.05in} \small (a) \hspace{1.35in} (b)\\[-1em]
 \caption{Examples of two different nearly data-consistent images $\mathbf{x} \in \Gamma_\varepsilon(\mathbf{A},\mathbf{b})$ that achieve the (a) upper- and (b) lower-bounds from Fig.~\ref{fig:bounds} for a specific column of voxels (indicated with red arrows). For simplicity, only the real parts of the images are shown (the imaginary parts of both images are  close to zero).}
 \label{fig:exemplar}
\end{figure}

An important thing to keep in mind is that the spatial maps of $L_i$ and $U_i$ shown in Fig.~\ref{fig:bounds} may happen to look like realistic MRI images, but they are not themselves elements of $\Gamma_\varepsilon(\mathbf{A},\mathbf{b})$.  In particular, these maps only provide entrywise bounds, and it is unlikely that these bounds could be achieved by all voxels simultaneously.  To illustrate this point,  Fig.~\ref{fig:exemplar} gives examples of two images belonging to $\Gamma_\varepsilon(\mathbf{A},\mathbf{b})$ that achieve the upper- or lower-bounds for specific choices of voxels (using the technique described in the proof of Thm.~\ref{thm:main} to identify the extremal cases).  

\begin{figure}[t]
\centering
\includegraphics[width=\columnwidth]{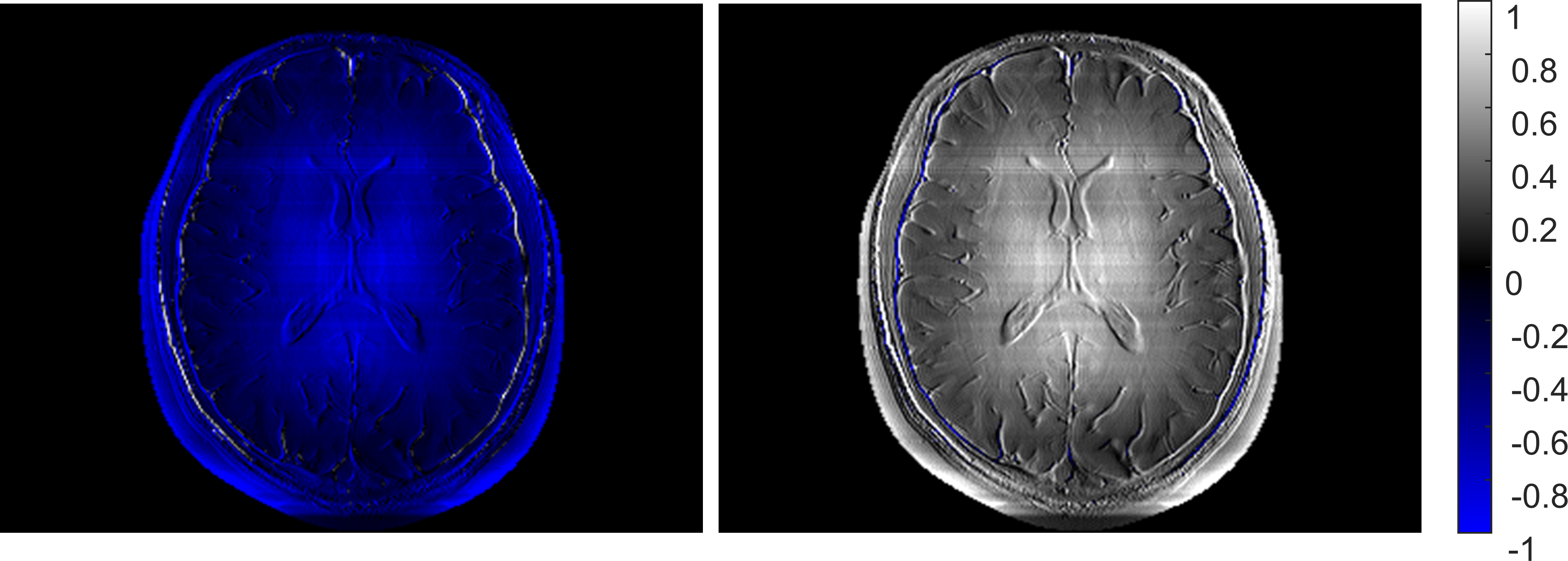}\\[-0.6em]
{\color{white}\,}\hspace{-0.3in} \small (a) $L_i$ \hspace{1.15in} (b) $U_i$ \\[-1em]
 \caption{Spatial maps of the (a)  lower-bounds $L_i$ and (b)  upper-bounds  $U_i$ for the difference between adjacent voxels, with $\mathbf{w} = \mathbf{e}_i - \mathbf{e}_j$.  }
 \label{fig:boundsD}
\end{figure}

As mentioned in Sec.~\ref{sec:theory}, we can also replace $\mathbf{e}_i$ in Thm.~\ref{thm:main} with a nonzero vector $\mathbf{w}\in\mathbb{R}^N$ to construct bounds on linear combinations of the entries of $\mathbf{x}$.  We illustrate this point by showing spatial maps of the upper- and lower-bounds on the difference between adjacent voxels ($\mathbf{w} = \mathbf{e}_i - \mathbf{e}_j$ as discussed in Sec.~\ref{sec:theory}, where $\mathbf{e}_i$ corresponds to the $i$th voxel position in the image and $\mathbf{e}_j$ corresponds to the voxel immediately to the right of the $i$th voxel) in Fig.~\ref{fig:boundsD}.  Interestingly, we observe that there are only a few voxels that are consistently larger or smaller than their neighbors (i.e., the lower-bounds and upper-bounds have consistent sign) within the set of nearly data-consistent solutions.

\begin{figure}[t]
\centering
\includegraphics[width=\columnwidth]{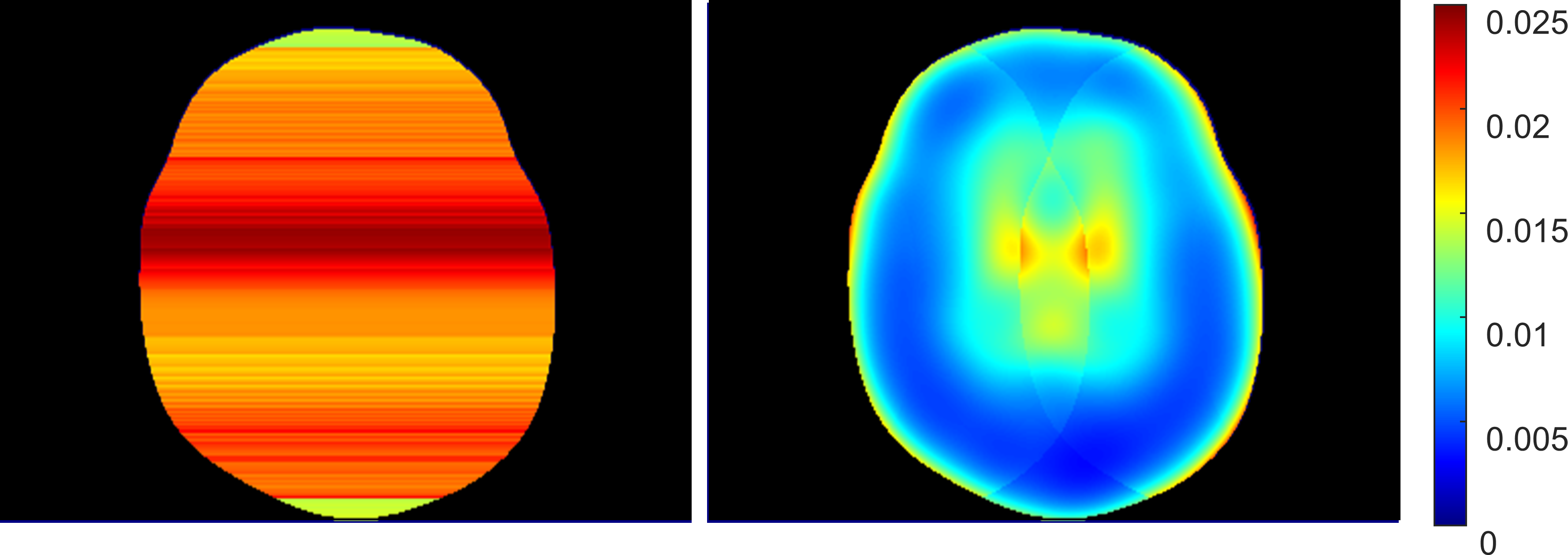}\\[-0.5em]
{\color{white}\,}\hspace{-0.2in} \small (a) $\sigma_N^{-1}(\mathbf{A})$ \hspace{0.8in} (b) $\|(\mathbf{A}^\dagger)^T\mathbf{e}_i\|_2$\\[-1em]
 \caption{Spatial maps of the spectral norm-type (a)  global  bounds $\sigma_N^{-1}(\mathbf{A})$ and (b)  entrywise bounds $\|(\mathbf{A}^\dagger)^T\mathbf{e}_i\|_2$. Note that because we use decoupling to solve for each image row independently, we have a different global bound for each image row.}
 \label{fig:spec}
\end{figure}

\begin{figure}[t]
\centering
\includegraphics[width=\columnwidth]{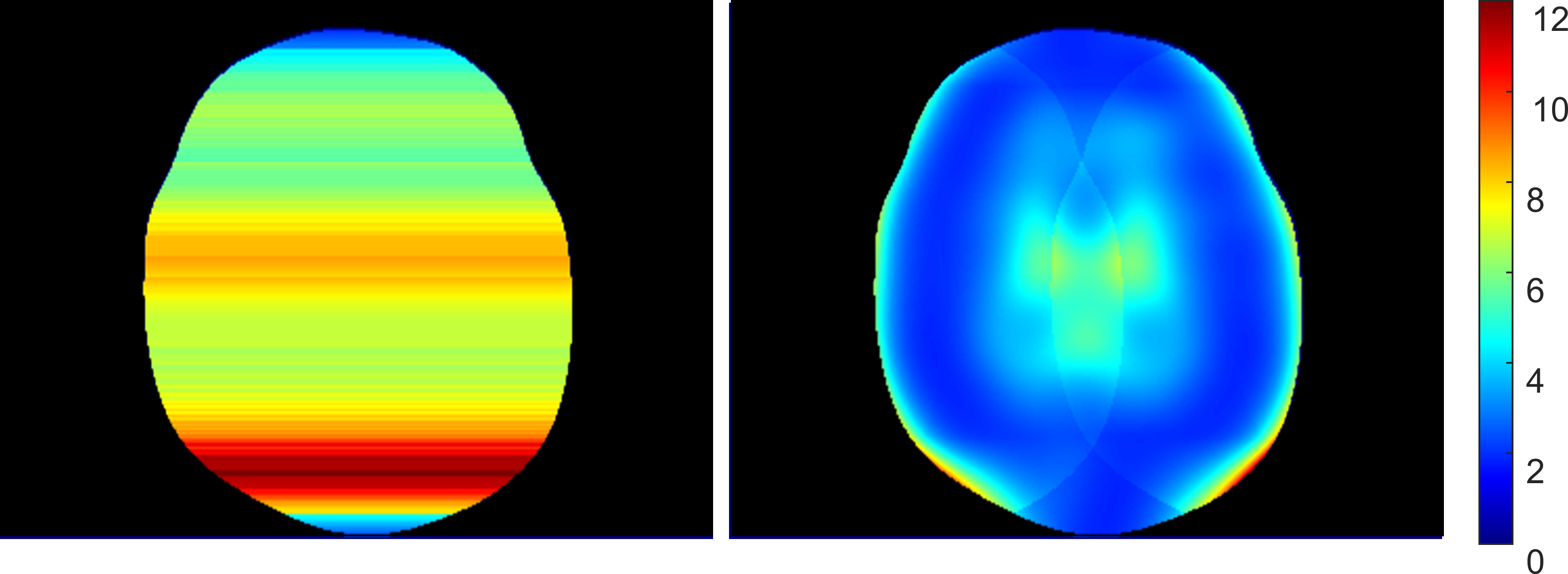}\\[-0.5em]
{\color{white}\,}\hspace{-0.4in} \small (a) $\kappa(\mathbf{A})$ \hspace{1.0in} (b) $\kappa_i(\mathbf{A})$\\[-1em]
 \caption{Spatial maps of the  (a)  traditional condition number $\kappa(\mathbf{A})$ and the (b)   entrywise condition number $\kappa_i(\mathbf{A})$. Note that because we use decoupling to solve for each image row independently, we have a different condition numbers for each image row.}
 \label{fig:cond}
\end{figure}

In our final set of illustrations, we compare the global  bounds from Eqs.~\eqref{eq:spec} and \eqref{eq:cond} against our new entrywise bounds from Cors.~\ref{cor3} and \ref{cor5}.  Specifically, Fig.~\ref{fig:spec} focuses on the spectral norm-type bounds from Eq.~\eqref{eq:spec} and Cor.~\ref{cor3}, while Fig.~\ref{fig:cond} focuses on the condition number bounds from Eq.~\eqref{eq:cond} and Cor.~\ref{cor5}. In order to show spatial maps of the global bounds in these figures, we have made use of the equivalence of norms  on finite dimensional spaces \cite{golub1996,bresler2008a}. Specifically, this equivalence implies that
\begin{equation}
 \|\mathbf{x}\|_\infty \leq \|\mathbf{x}\|_2 \leq \sqrt{N} \|\mathbf{x}\|_\infty
\end{equation}
for arbitrary $\mathbf{x} \in \mathbb{R}^N$, where $\|\cdot\|_\infty$ is the standard infinity norm. This relationship immediately allows a global bound in the form of Eq.~\eqref{eq:global} to be converted into a bound on the individual elements of $\mathbf{x}$:
\begin{equation}
 \|\hat{\mathbf{x}} - \mathbf{x}^*\|_\infty \leq C\|\mathbf{n}\|_2,
\end{equation}
which allows us to depict spatial maps of the global bounds.

As can be seen from Figs.~\ref{fig:spec} and \ref{fig:cond}, our proposed entrywise bounds are usually better and never worse than the global bounds, consistent with our theoretical expectations.  In addition, the  entrywise characterizations both do a good job of capturing the fact that the values of certain voxels can be substantially less sensitive to noise perturbations than others, while the traditional global characterizations lack this nuance and are more pessimistic.

\section{Discussion}\label{sec:disc}
\subsection{Efficient Computations for Large Problems}
The new theoretical characterizations presented in this paper all depend on the ability to compute the numerical value of  $\left\| \left(\mathbf{A}^\dagger\right)^T\mathbf{e}_i\right\|_2$.   When the problem size is small enough, it is possible to explicitly calculate the SVD of $\mathbf{A}$, at which point this calculation becomes trivial.  However, in many practical scenarios, the matrix $\mathbf{A}$ is too big to be stored in memory, and it is necessary to work with functions that compute matrix-vector multiplications with $\mathbf{A}$ or $\mathbf{A}^T$ rather than working directly with the matrix $\mathbf{A}$.  

In situations like this, there exist well-established iterative procedures that allow computation of the quantity $\mathbf{A}^\dagger\mathbf{m}$ for arbitrary vectors $\mathbf{m} \in \mathbb{R}^M$ and arbitrary matrices $\mathbf{A}$.  For example, starting from zero initialization $\mathbf{x}^{(0)} = \mathbf{0}$, the Landweber iteration \cite{landweber1951,bertero1998,hansen1997,vogel2002}\footnote{Other iterative algorithms like the conjugate gradient method \cite{hestenes1952} or LSQR \cite{paige1982} could also be employed to determine $\mathbf{A}^\dagger\mathbf{m}$ for arbitrary vectors $\mathbf{m}$. These methods will generally converge faster than Landweber iteration, although it can be important to watch out for numerical stagnation issues that may prevent these algorithms from converging to the desired value of $\mathbf{A}^\dagger\mathbf{m}$.}
\begin{equation}
 \mathbf{x}^{(k)} = \mathbf{x}^{(k-1)} - \tau \mathbf{A}^T(\mathbf{A}\mathbf{x}^{(k-1)} - \mathbf{m})
\end{equation}
will satisfy $\mathbf{x}^{(k)} \rightarrow \mathbf{A}^\dagger\mathbf{m}$ as $k\rightarrow \infty$ as long as $0<\tau <2/\sigma_1^2(\mathbf{A})$.  Furthermore, if the value of $\sigma_r(\mathbf{A})$ is known where $r$ is the rank of $\mathbf{A}$, then it is also possible to use this value to precompute the number of iterations $k$ that are needed so that $\mathbf{x}^{(k)}$ approximates $\mathbf{A}^\dagger\mathbf{m}$ within a prescribed level of accuracy.\footnote{While it is common to truncate the iterations early  to take advantage of semiconvergence phenomena when using Landweber iteration to solve inverse problems \cite{hansen1997,bertero1998,vogel2002}, it is important to not use early-stopping criteria when computing our theoretical bounds, as this can lead to an underestimation of the ambiguity in the solution to the inverse problem.}  This type of approach allows  for simple computation of $\mathbf{A}^\dagger \mathbf{e}_i$.

However, this type of approach quickly becomes computationally burdensome if we wish to compute $\left\| \left(\mathbf{A}^\dagger\right)^T\mathbf{e}_i\right\|_2$ for a large number of different $i$ values.  Fortunately, it is also possible to use stochastic methods to approximately compute $\left\| \left(\mathbf{A}^\dagger\right)^T\mathbf{e}_i\right\|_2$ for all $i$ simultaneously.  In particular, let $\mathbf{z}_s \in \mathbb{R}^M$ for $s=1,\ldots,S$ be a collection of independent and identically distributed random vectors with zero mean and covariance matrix equal to $\mathbf{I}_M$, and let 
\begin{equation}
 \mathbf{g}_S \triangleq \frac{1}{S}\sum_{s=1}^S\left|\mathbf{A}^\dagger \mathbf{z}_s\right|^2,
\end{equation}
where our notation assumes that the magnitude-squaring operation is applied separately to each entry of the vector.  Then it is not hard to show that the expected value of $i$th entry of $\mathbf{g}_S$ is equal to $\left\| \left(\mathbf{A}^\dagger\right)^T\mathbf{e}_i\right\|_2^2$, and that $\mathbf{g}_S$ converges to its expected value  in the limit as $S\rightarrow \infty$.    This allows approximate calculation of $\left\| \left(\mathbf{A}^\dagger\right)^T\mathbf{e}_i\right\|_2$ for all $N$ possible values of $i$ simultaneously, but only requiring the computation of $\mathbf{A}^\dagger\mathbf{z}_s$ for $s=1,\ldots,S$.  If $S \ll N$, then this can represent a massive improvement in computational complexity over directly calculating $\|(\mathbf{A}^\dagger)^T\mathbf{e}_i\|_2$ for each $i=1,\ldots,N$.  Notably, this approach is quite similar to stochastic techniques designed to efficiently estimate the diagonal entries of spatially-varying covariance matrices \cite{robson2008}.

\subsection{Relationship to Cram\'{e}r-Rao bounds}

The theoretical bounds we obtained in this paper were derived without making any assumptions about the statistical characteristics of $\mathbf{n}$, which can be useful when the distribution of $\mathbf{n}$ is unknown.  Interestingly, if we make stronger assumptions about the statistical characteristics of $\mathbf{n}$, then we can establish direct links between the traditional Cram\'{e}r-Rao bound from estimation theory \cite{kay1993} and a quantity that appears prominently in our theoretical bounds, namely $\|(\mathbf{A}^\dagger)^T\mathbf{e}_i\|_2$.\footnote{Readers from the MRI community may benefit from knowing that the Cram\'{e}r-Rao bound is highly related to the  \emph{g-factor} from the MRI literature \cite{pruessmann1999,robson2008}.  Specifically, the g-factor can simply be viewed as the normalized voxelwise ratio between the Cram\'{e}r-Rao bounds for subsampled and fully-sampled acquisitions. }  

To derive Cram\'{e}r-Rao bounds, we will now make the assumption that $\mathbf{n}$ is a zero-mean Gaussian random vector with covariance matrix $\mathbf{I}_M$.  In this case, assuming that $\mathbf{e}_i \perp \mathcal{N}(\mathbf{A})$, it can be shown that the entrywise Fisher information corresponding to the quantity $x_i = \mathbf{e}_i^T\mathbf{x}$  is equal to $(\mathbf{e}_i^T(\mathbf{A}^T\mathbf{A})^\dagger \mathbf{e}_i)^{-1}$ \cite{pukelsheim1993}. This means that the entrywise Cram\'{e}r-Rao bound for $x_i$  (which is this inverse of the entrywise Fisher information, as well as a lower bound on the variance of any unbiased estimator of $x_i$ \cite{kay1993}) is equal to $\mathbf{e}_i^T(\mathbf{A}^T\mathbf{A})^\dagger \mathbf{e}_i = \|(\mathbf{A}^\dagger)^T \mathbf{e}_i\|_2^2$.  This establishes a clear parallel between Cram\'{e}r-Rao bounds and our theoretical characterization.

This relationship is potentially interesting because minimization of the Cram\'{e}r-Rao bound is a common objective in methods to design optimally-efficient $\mathbf{A}$ matrices \cite{pukelsheim1993}. For example, this type of Cram\'{e}r-Rao optimization approach  has been explored  to optimize data acquisition protocols in MRI applications with linear  data acquisition models \cite{reeves1999,xu2005,haldar2018}.  While this approach has historically been motivated by  statistical variance-minimization arguments, our theoretical results demonstrate that this approach can also be justified based on deterministic principles, without requiring strong assumptions about the nature of the noise $\mathbf{n}$.  Specifically, minimizing the Cram\'{e}r-Rao bound for $x_i$ under white Gaussian noise assumptions implicitly has the same effect as minimizing the length of the interval $[L_i, U_i]$ for $x_i$ under near data-consistency constraints, regardless of the actual distribution of $\mathbf{n}$!  

It should be noted that we have only established a relationship between our new entrywise bounds  and entrywise Cram\'{e}r-Rao bounds  for linear models in the form of Eq.~\eqref{eq:model}.  There are also many applications where nonlinear data acquisition procedures are optimized by minimizing Cram\'{e}r-Rao bounds (e.g., a few MRI examples include \cite{jones1996,marseille1994,cavassila2001,pineda2005,alexander2008,haldar2009a,denaeyer2011,zhao2017}).  Based on our experience with the linear case, we suspect that it may be possible to rigorously justify the goodness of such design approaches for nonlinear models without making strong statistical assumptions, and believe that this could be an interesting topic for future research.

\subsection{Extensions to Constrained Reconstruction?}
One of the features of our  theory is that our bounds rely entirely on near data-consistency constraints $\Gamma_\varepsilon(\mathbf{A},\mathbf{b})$, and can be agnostic to the specific estimation methods that are used to obtain $\hat{\mathbf{x}}$ or any additional prior information that we may have about $\mathbf{x}^*$.  In practice, modern estimation methods often rely on more than just near data-consistency, and also impose other forms of constraints when solving an inverse problem.  For example, in modern MRI reconstruction, it is common to look for reconstructed images that are not only nearly-data consistent, but which also obey sparsity constraints \cite{lustig2007,block2007,trzasko2009,jung2009a,haldar2010a}, low-rank constraints \cite{liang2007,haldar2010,goud2011,zhao2012,trzasko2011a,otazo2015},  autoregressive/structured low-rank constraints \cite{haldar2020,shin2014,haldar2013b,jin2015a,ongie2016}, or even manifold constraints that could be learned by applying machine learning methods to large databases of previous images \cite{knoll2020,liang2020,sandino2020}.  (See also \cite{liang1992, haldar2022 } for additional historical context).  

These kinds of additional constraints are all expected to reduce ambiguity in the solution to the inverse problem.  As such, rather than just understanding elementwise ambiguity for the set of nearly data-consistent solutions $\mathbf{x} \in \Gamma_\varepsilon(\mathbf{A},\mathbf{b})$, it may  be even better to be able to identify elementwise ambiguity for solutions that are both nearly data-consistent and obey other constraints, i.e., $\mathbf{x} \in \Gamma_\varepsilon(\mathbf{A},\mathbf{b}) \cap \mathcal{K}$, where $\mathcal{K}$ is a  set that might represent a sparsity constraint, a low-rank constraint, a manifold constraint, etc.  

This is a difficult problem and a rigorous investigation of the interplay between $\Gamma_\varepsilon(\mathbf{A},\mathbf{b})$ and different types of complicated constraint sets $\mathcal{K}$ would go far beyond the scope of this paper, though we expect this to be a promising direction for future exploration.  We especially believe that developing interval bounds of the form $x_i \in [L_i, U_i]$ for $\forall \mathbf{x} \in  \Gamma_\varepsilon(\mathbf{A},\mathbf{b}) \cap \mathcal{K}$ may be important for building trust in machine-learning methods designed to solve inverse problems, given the potential for instability and limited resolving power that has been observed with some of these kinds of methods \cite{antun2020,chan2021}.

\bibliographystyle{IEEEtran}
\bibliography{./reference}

\begin{IEEEbiographynophoto}{Justin P. Haldar} (Senior Member, IEEE) received the Ph.D. in Electrical and Computer Engineering from the University of Illinois at Urbana-Champaign in 2011.  He is now an Associate Professor in the Signal and Image Processing Institute and the Ming Hsieh Department of Electrical and Computer Engineering at the University of Southern California. His research interests include inverse problems, signal processing, computational imaging, and applications in biomedical imaging (especially magnetic resonance imaging). His research has been recognized with honors such as the NSF CAREER award, the IEEE ISBI best paper award, and the IEEE EMBC first-place student paper award, among others. He is an Associate Editor for the IEEE Transactions on Medical Imaging (2014-Present) and a Senior Area Editor for the IEEE Transactions on Computational Imaging (2021-Present; previously Associate Editor from 2018-2021), and has twice received (in 2019 and 2022) the Outstanding Editorial Board Award from the IEEE Signal Processing Society. He has also been active in the IEEE Signal Processing Society's Technical Committees on Computational Imaging (Chair Elect, 2024-2025; Vice Chair, 2023; Advisory Member, 2022; Member, 2016-2021) and Bio Imaging and Signal Processing (Member, 2014-2019; Associate Member, 2011-2013).
\end{IEEEbiographynophoto}

\end{document}